\documentclass[twocolumn,trackchanges]{aastex701}

\shorttitle{Baryonic Content in DM Halos}
\shortauthors{Contini et al.}

\begin{document}

\title{Roles of Supernova and Active Galactic Nucleus Feedback in Shaping the Baryonic Content in a Wide Range of Dark Matter Halo Masses}

\author{Emanuele Contini}
\altaffiliation{Yonsei University}
\affiliation{Department of Astronomy and Yonsei University Observatory, Yonsei University, 50 Yonsei-ro, Seodaemun-gu, Seoul 03722, Republic of Korea}
\email[show]{emanuele.contini82@google.com}

\author{Changjo Seo}
\altaffiliation{Yonsei University}
\affiliation{Department of Astronomy and Yonsei University Observatory, Yonsei University, 50 Yonsei-ro, Seodaemun-gu, Seoul 03722, Republic of Korea}
\email{emanuele.contini82@gmail.com}

\author{Jinsu Rhee}
\altaffiliation{Yonsei University}
\affiliation{Department of Astronomy and Yonsei University Observatory, Yonsei University, 50 Yonsei-ro, Seodaemun-gu, Seoul 03722, Republic of Korea}
\affiliation{Institut d’Astrophysique de Paris, Sorbonne Université, CNRS, UMR 7095, 98 bis bd Arago, 75014 Paris, France}
\email{emanuele.contini82@google.com}  

\author{Seyoung Jeon}
\altaffiliation{Yonsei University}
\affiliation{Department of Astronomy and Yonsei University Observatory, Yonsei University, 50 Yonsei-ro, Seodaemun-gu, Seoul 03722, Republic of Korea}
\email{emanuele.contini82@google.com}  

\author{Sukyoung K. Yi}
\altaffiliation{Yonsei University}
\affiliation{Department of Astronomy and Yonsei University Observatory, Yonsei University, 50 Yonsei-ro, Seodaemun-gu, Seoul 03722, Republic of Korea}
\email[show]{yi@yonsei.ac.kr}

\begin{abstract}
We build upon \texttt{FEGA25} \citep{contini2025a}, a previously introduced semi-analytic model (SAM) for galaxy formation and evolution, by focusing on its enhanced treatment of supernova (SN) and active galactic nucleus (AGN) feedback mechanisms.
In addition to the traditional AGN feedback mode—negative (suppressing cooling) and the new positive mode (triggering star formation)—we introduce two distinct implementations of a third mode: the ejection of hot gas beyond the virial radius (\texttt{AGNeject1} and \texttt{AGNeject2}). This component addresses a longstanding issue in SAMs and hydrodynamical simulations: the overestimation of hot gas fractions in low- and intermediate-mass halos.
\texttt{FEGA25} is calibrated via MCMC using a suite of cosmological N-body simulations (YS50HR, YS200, YS300) and a comprehensive set of observed stellar mass functions across a wide redshift range. We find that SN feedback dominates gas ejection in halos with $\log M_{\rm halo} \lesssim 12$, while AGN feedback becomes increasingly important at higher halo masses. The \texttt{AGNeject2} model, which activates primarily at late times ($z < 1$), reproduces a characteristic ``cavity" (sort of U-shape) in the baryon fraction at $z = 0$, similar to trends observed in simulations like SIMBA and IllustrisTNG. Conversely, \texttt{AGNeject1} yields a smoother, redshift-independent evolution.
Both models preserve the stellar and cold gas components and successfully reproduce the stellar-to-halo mass relation up to $z = 3$. Our results emphasize that a physically motivated AGN-driven mechanism capable of selectively removing hot gas is essential to model the baryon cycle accurately, particularly in the intermediate halo mass regime.

\end{abstract}



\section{Introduction}
\label{sec:intro}

In the current $\Lambda$CDM framework, galaxies form within dark matter (DM) halos that assemble hierarchically through mergers and smooth accretion \citep{white1978,springel2005}. While the growth of DM structures is well described by gravitational dynamics, the observable properties of galaxies depend critically on baryonic processes—namely gas cooling, star formation, chemical enrichment, and feedback. Among these, feedback from supernovae (SN) and active galactic nuclei (AGN) is widely regarded as essential for regulating star formation, quenching massive galaxies, and explaining the observed galaxy stellar mass function \citep{guo2013,somerville2015,naab2017,contini2019b,contini2020,vogelsberger2020,henriques2020}.

Yet, despite the inclusion of increasingly sophisticated feedback prescriptions, models still struggle to simultaneously match key observables such as the cosmic star formation rate density \citep{madau2014}, the stellar-to-halo mass relation \citep{moster2013,behroozi2019}, and the distribution of hot and cold baryons across halo mass and redshift \citep{chiu2016,eckert2021,popesso2024}. This tension is particularly severe in the group mass regime ($\log M_{\rm halo} \sim 12.5$–13.5), where feedback must be powerful enough to eject baryons, yet gentle enough not to alter the stellar and cold gas components.

Large-scale hydrodynamic simulations such as IllustrisTNG \citep{pillepich2018,nelson2019}, SIMBA \citep{dave2019}, EAGLE \citep{schaye2015}, MAGNETICUM \citep{dolag2016}, and more recently FLAMINGO \citep{schaye2023} and MillenniumTNG \citep{pakmor2023}, have offered valuable insights into feedback-regulated galaxy evolution. However, they often yield divergent predictions for baryon fractions and hot gas content, reflecting the sensitivity of such quantities to the underlying physics. For instance, while IllustrisTNG and SIMBA predict a baryon ``cavity”—a depression in the baryon fraction at intermediate halo masses—other simulations such as MAGNETICUM show a smoother, power-law trend more consistent with X-ray and SZ observations \citep{hadzhiyska2024,popesso2024}.

Recent comparative analysis \citep{ayromlou2023,salcido2023,dev2024} have further demonstrated that baryon retention is highly dependent on the efficiency and coupling of feedback modes. These studies reinforce the need for physically motivated models that can selectively eject hot gas—especially in group-scale halos—without modifying the stellar and cold gas reservoirs \citep{eckert2021,dev2024,popesso2024}.

In this context, semi-analytic models (SAMs) remain an indispensable tool. Their modular structure and computational efficiency allow for rapid exploration of parameter space and clearer isolation of physical mechanisms. Modern SAMs such as those developed by \citet{lagos2024} and \citet{contini2025a} have adopted increasingly sophisticated AGN feedback prescriptions, including both radio-mode heating and direct ejection of hot gas. Notably, the SHARK2 model \citep{lagos2024} includes a formulation similar in spirit to one of our schemes.

Building on this foundation, we have developed the \texttt{FEGA} series \citep{contini2024d,contini2025a}, a SAM calibrated against a wide range of observational constraints. In particular, \texttt{FEGA25} \citep{contini2025a} introduced two key advancements: a physically motivated positive AGN feedback mode (triggering star formation under residual cooling), and a novel mechanism for ejecting hot gas beyond the virial radius. The latter is motivated by the need to reconcile models with observed hot gas fractions in low- and intermediate-mass halos, where supernova feedback alone appears insufficient \citep{henriques2013,hirschmann2016,ayromlou2023}.

In this work, we investigate the impact of two distinct implementations of this hot gas ejection mode—\texttt{AGNeject1}, which scales with black hole accretion and halo virial velocity, and \texttt{AGNeject2}, which uses the surplus AGN energy beyond that needed to suppress cooling to drive ejection. We calibrate both models using cosmological DM–only simulations and an extensive set of observed stellar mass functions from $z=3$ to the present \citep{marchesini2009,marchesini2010,ilbert2013,muzzin2013,tomczak2014,bernardi2018}.

We focus on three main goals: (i) to quantify the relative contributions of SN and AGN feedback in ejecting baryons beyond the virial radius; (ii) to assess the evolution of baryon and hot gas fractions as a function of halo mass and redshift; and (iii) to compare our predictions with hydrodynamic simulations and observational estimates, particularly those from recent X-ray and SZ studies \citep{chiu2016,eckert2021,angelinelli2023,popesso2024}.

Our analysis offers new insights into the physical mechanisms driving baryon depletion in halos and provides a framework for interpreting upcoming observational constraints. In particular, we argue that the inclusion of a physically motivated AGN-driven hot gas ejection channel is essential for reproducing both the amplitude and shape of baryon and hot gas fraction trends across halo mass—especially in the transition regime between SN- and AGN-dominated feedback.

In Section~\ref{sec:model}, we present the key features of \texttt{FEGA25}, highlighting the implementation of its AGN feedback modes and, in particular, the two alternative schemes for hot gas ejection—\texttt{AGNeject1} and \texttt{AGNeject2}. We describe the set of cosmological N-body simulations used to generate DM halo merger trees, the observational constraints employed for calibration, and the methodology adopted for parameter tuning via Markov Chain Monte Carlo (MCMC). Moreover, part of the analysis will be complemented by results from our in-house NewCluster simulation, which will be briefly described.

In Section~\ref{sec:results}, we analyze the resulting galaxy catalogs and compare the predicted baryonic content—total baryon fraction, hot gas fraction, and stellar mass—with a comprehensive set of observations and hydrodynamic simulations. We quantify the efficiency of SN and AGN feedback in removing gas beyond the virial radius, assess the emergence of baryon depletion features such as the cavity, and evaluate the redshift evolution of these effects.

Finally, in Section~\ref{sec:conclusion}, we summarize our main findings and discuss their implications for galaxy formation theory and future observational efforts. Stellar masses are computed by adopting a \citet{chabrier2003} initial mass function, and unless otherwise specified, all masses are h-corrected.

\section[]{Methods}
\label{sec:model}

To achieve our goals, we employ the state-of-the-art semi-analytic model \texttt{FEGA25} (Formation and Evolution of GAlaxies \citealt{contini2025a}, hereafter C25), applied to a suite of three DM–only cosmological simulations. Additionally, we gain further insight from our high-resolution zoom-in simulation \texttt{NewCluster} (\citealt{han2025}). Below, we briefly outline the key features of our SAM, including the main physical prescriptions relevant to this study and, most notably, the new implementation of an alternative AGN-driven hot gas ejection mode, which represents a key innovation of this work. We also summarize the main characteristics of the \texttt{NewCluster} simulation.
It is important to emphasize that in the analysis presented in Section~\ref{sec:results}, we make limited use of \texttt{NewCluster} predictions, as a dedicated and more comprehensive study based on this simulation is currently in preparation (Seo C., et al.).

\subsection{\texttt{FEGA25}}

\texttt{FEGA25} is an updated version of the SAM originally introduced in \citealt{contini2024d} (hereafter C24), featuring two major advancements: (i) a more sophisticated treatment of star formation that follows the Extended Kennicutt–Schmidt relation \citep{shi2011}, and (ii) a novel implementation of positive AGN feedback, a mode that enhances star formation rather than suppressing it.

The main new feature introduced in \texttt{FEGA25} is a third mode of AGN feedback, specifically designed to eject hot gas beyond the virial radius of halos. This mechanism, referred to as the hot gas ejection mode, adds to the existing suite of feedback processes already present in the SAM. In this framework, gas can be expelled from halos via two channels: SN feedback and AGN-driven ejection.

The three AGN feedback (radio) modes, along with SN feedback, are the key physical processes governing baryon cycling in our model and are central to the present study. Below, we describe each of them in detail.

Finally, as mentioned above, we introduce an alternative implementation of the hot gas ejection mode. For clarity, we refer to the original version as \texttt{AGNeject1}, and to the updated one as \texttt{AGNeject2}.

\subsubsection{SN feedback}
SN feedback, resulting from the explosive deaths of massive stars, is a fundamental process in galaxy formation and is central to the present model. This mechanism is particularly effective in regulating star formation in low-mass galaxies, shaping the low-mass end of the stellar mass function (SMF) and luminosity function (e.g., \citealt{guo2010, henriques2013}) by limiting stellar mass growth in dwarfs. The energy released—both radiative and kinetic—can substantially impact the interstellar medium, leading to its heating and ionization. In some cases, this energetic input is sufficient to drive gas completely out of the galactic halo, making SN feedback a critical component in virtually all SAMs of galaxy formation. Historically, SN feedback has been maximized in such models because it has been essential to reproducing the observed evolution of the SMF’s low-mass slope (e.g., \citealt{guo2011, guo2013, henriques2013, henriques2015, hirschmann2016, contini2017, henriques2020}).

In \texttt{FEGA25}, SN feedback is implemented such that cold gas in the galactic disk is heated and transferred to the hot gas phase, and part of the hot gas is further expelled into an external reservoir. This approach follows the formulation of \citet{guo2011}, though we adopt different parameters, following \citet{hirschmann2016}, in which the feedback weakens over cosmic time.

In this framework, the amount of cold gas reheated by SN explosions and moved to the hot phase is given by:

\begin{equation}
\delta M_{\rm{reh}} = \epsilon_{\rm{reh}} \cdot \left[0.5+(1+z)^3 \left(\frac{V_{\rm{max}}}{V_{\rm{reh}}}\right) ^{-\beta_{\rm{reh}}}\right]\cdot \delta M_* ,
\end{equation}
where $\epsilon_{\rm{reh}}=0.7$, $V_{\rm{reh}}=70\, \rm{km/s}$, $\beta_{\rm{reh}}=3.5$, and $\delta M_*$ represents the mass of newly formed stars. The inclusion of the variable $\epsilon_{\rm{reh}}$ by \cite{guo2011} allows for enhanced ejection efficiencies in dwarf galaxies. The total energy injected into the disk and halo gas is:
\begin{equation}
\delta E_{\rm{SN}} = \eta_{\rm{ej}} \cdot \left[0.5+(1+z)^3 \left(\frac{V_{\rm{max}}}{V_{\rm{ej}}}\right)^{-\beta_{\rm{ej}}} \right]\cdot \delta M_* \cdot 0.5V_{\rm{SN}}^2,
\end{equation}
where $\eta_{\rm{ej}}=0.15$, $V_{\rm{ej}}=70\, \rm{km/s}$, $\beta_{\rm{ej}}=3.5$, and $0.5V_{\rm{SN}}^2$ represents the average kinetic energy of SN ejecta per unit mass of stars formed, calculated based on $V_{\rm{SN}}=630$ km/s (\citealt{croton2006}). From this energy amount, the ejected gas (including metals) from the halo is:
\begin{equation}
\delta M_{\rm{ej}} = \frac{\delta E_{\rm{SN}}-0.5\delta M_{\rm{reh}}V_{200}^2}{0.5V_{200}^2}.
\end{equation}
The ejected material populates the ejecta reservoir and may be reincorporated at later stages (see C24 for further details).

\subsubsection{Negative and Positive AGN feedback}

The radio mode of AGN feedback in our model follows the original prescription of \cite{croton2006}. In this mode, hot gas accretes onto the central black hole (BH), releasing mechanical energy that heats the surrounding halo gas. The accretion rate is given by:

\begin{equation}\label{eqn:radiomode}
\dot{M}_{\rm{BH}} = \kappa_{\rm{AGN}} \left(\frac{f_{\rm{hot}}}{0.1}\right)\left(\frac{V_{200}}{200\,  \rm{km/s}}\right)^3 \left(\frac{M_{\rm{BH}}}{10^8 \, M_{\odot}/h}\right)
\end{equation}
where $f_{\rm{hot}}$ is the hot gas fraction (relative to halo mass), and $\kappa_{\rm{AGN}}$ is a free parameter tuned during calibration to regulate accretion efficiency.

The energy released through this process is:

\begin{equation}\label{eqn:energy}
\dot{E}_{\rm{radio}} = \eta_{\rm{rad}} \dot{M}_{\rm{BH}}c^2 ,
\end{equation}
with $\eta_{\rm{rad}} = 0.1$ and $c$ being the speed of light. This energy injection offsets gas cooling, leading to a revised cooling rate:

\begin{equation}\label{eqn:mcool}
\dot{M}_{\rm{cool,new}} = \dot{M}_{\rm{cool}}-2\frac{\dot{E}_{\rm{radio}}}{V_{200}^2},
\end{equation}

Depending on the AGN power, cooling may be partially or fully suppressed—concluding the negative AGN feedback phase. Traditionally, SAMs channel the net cooled mass into the cold gas reservoir for star formation. However, in \texttt{FEGA25}, this component undergoes an additional feedback process: the positive mode.

Introduced in C24, the positive AGN feedback simulates a secondary starburst triggered by AGN-driven pressure. This effect, while supported by both observational (e.g., \citealt{cresci2015a, mahoro2017, joseph2022}) and theoretical studies (\citealt{gaibler2012, silk2013, mukherjee2018}), was initially modeled in a simplified way to test its impact, but introducing new free parameters.

In \texttt{FEGA25}, we maintain this minimal approach. The positive feedback mode uses the same parameters as the negative mode and is only triggered if some residual cooling occurs ($\dot{M}_{\rm{cool,new}} > 0$). The resulting star formation rate from this mode is given by:

\begin{equation}\label{eqn:newpAGN}
\dot{M}_* = \left(\frac{\delta M_{\rm{BH}}}{M_{\rm{BH}}}\right) \left(\frac{\dot{M}_{\rm{cool,new}}}{10^{8}M_{\odot}/h}\right),
\end{equation}
where $\delta M_{\rm{BH}}$ is the BH mass growth due to accretion (Eq.~\ref{eqn:radiomode}).

This new formulation (implemented in C25) improves on the earlier version (the original one in C24) by tying the efficiency of the positive mode directly to BH accretion, which is itself regulated by $\kappa_{\rm{AGN}}$. This ensures that both AGN feedback modes are consistently governed by a single parameter. It also introduces a natural coupling between them: the positive mode becomes prominent when the negative one is weak (e.g., at high redshift), and diminishes as the negative feedback grows stronger over time.

\subsubsection{\texttt{AGNeject1} and \texttt{AGNeject2}}

In C25, we outlined a longstanding difficulty in SAMs (but also in hydrodynamic simulations): their failure to accurately reproduce the fraction of hot gas in halos, along with a proposed strategy to resolve it. This solution requires a mechanism capable of removing part of the hot gas beyond the virial radius, while leaving cooling and star formation processes unaffected. That is, the mechanism should extract hot gas without altering the cold gas or stellar components. To achieve this, we introduced a novel AGN feedback channel in \texttt{FEGA25}, referred to as the hot gas ejection mode, \texttt{AGNeject1}.

Inspired by the same rationale used in C24 to isolate the positive feedback effect, we postulate that such a mechanism is physically feasible and posit that the AGN can remove a portion of the hot gas, transferring it to an “ejected” reservoir. In our SAM, the amount of expelled hot gas is assumed to scale with the efficiency of the two other AGN feedback modes, through the parameter $\kappa_{\rm{AGN}}$, and to be modulated by the virial velocity of the host halo. This ensures that expelling gas becomes increasingly difficult in more massive systems. The mass of hot gas expelled in \texttt{FEGA25} is computed via:

\begin{equation}\label{eqn:ejection}
M_{\rm{ejected}} = \left(\frac{\delta M_{\rm{BH}}}{M_{\rm{BH}}}\right) \left(\frac{M_{\rm{hot}}}{10^{8}M_{\odot}/h}\right)\left(1-\frac{V_{\rm{200}}}{V_{\rm{scale}}}\right),
\end{equation}
where $V_{\rm{scale}}$ is a tunable parameter determined during model calibration, and it sets the threshold at which this ejection becomes efficient \footnote{This prescription remains valid even when $\dot{M}_{\rm{cool}}>2\frac{\dot{E}_{\rm{radio}}}{V_{200}^2}$, i.e., when the AGN energy input is insufficient to fully suppress gas cooling.}. High values of $V_{\rm{scale}}$ make gas ejection more effective in massive halos, whereas low values diminish efficiency in low-mass halos. The computed ejected mass, $M_{\rm{ejected}}$, is stored in an external reservoir but remains eligible for future reincorporation, consistent with the treatment described in C24. Naturally, when $V_{200} > V_{\rm{scale}}$, the ejected mass is set to zero.

As discussed earlier, negative AGN feedback can, in certain regimes, deliver enough energy to entirely offset gas cooling, effectively reheating gas to the virial temperature. However, in cases where the energy released by the AGN surpasses what is needed to quench the cooling flow, it is plausible that this surplus energy could drive gas out of the halo. We propose here, and this is \texttt{AGNeject2}, that the excess energy, beyond what is required to halt cooling, may be responsible for ejecting gas outside the virial boundary.

Analogously to the implementation in \texttt{AGNeject1}, we assume that gas ejection becomes progressively more challenging in more massive halos—specifically, it is governed by the virial velocity of the host halo. Consequently, in cases where $\dot{M}_{\rm{cool}} < 2\frac{\dot{E}_{\rm{radio}}}{V_{200}^2}$, the surplus energy is assumed to expel a portion of the excess gas mass, $$M_{\rm{excess}}=\left(2\frac{\dot{E}_{\rm{radio}}}{V_{200}^2}-\dot{M}_{\rm{cool}}\right)\cdot \rm dt ,$$ which would otherwise have been reheated to the virial temperature. The amount of ejected mass is computed using the following expression:

\begin{equation}\label{eqn:ejection2}
M_{\rm{ejected}} = M_{\rm{excess}}\left(1-\frac{V_{\rm{200}}}{V_{\rm{scale}}}\right),
\end{equation}
with $V_{\rm{scale}}$ that must be determined with a separate calibration.

In summary, the full AGN feedback framework implemented in \texttt{FEGA25} consists of three components: the negative mode, the positive mode, and the hot gas ejection mode, either \texttt{AGNeject1} or \texttt{AGNeject2}. Compared to the version presented in C24, \texttt{FEGA25} introduces a more physically motivated description of the positive feedback (see Equation~\ref{eqn:newpAGN}) and adds a new prescription for the ejection of hot gas (Equations~\ref{eqn:ejection} or ~\ref{eqn:ejection2}). As a result, \texttt{FEGA25} stands out as the only SAM that incorporates both a treatment of AGN-driven positive feedback and a mechanism for the ejection of hot gas. It must be noted, however, that an implementation similar to \texttt{AGNeject2} is incorporated in SHARK2 SAM \citep{lagos2024}.

Our objective is to investigate the differences between the two hot gas ejection modes and, more importantly, to assess the relative roles of SN and AGN feedback in expelling hot gas beyond the virial radius across a wide range of halo masses.

\subsubsection{Set of Simulations and Calibration}

Calibrating a SAM is a crucial step, as the physical prescriptions governing galaxy formation involve many parameters that are often poorly constrained. We calibrate \texttt{FEGA25} using the same MCMC approach as in C25. For a broader discussion of this method, including its advantages and limitations, we refer to C24, \citet{henriques2020}, and references therein.

The calibration relies on the same compilation of observed SMFs used in C24 and C25, spanning $z=3$ to $z=0$, with data from \citet{marchesini2009,marchesini2010,ilbert2010,sanchez2012,muzzin2013,ilbert2013,tomczak2014} at high redshift, and \citet{baldry2008,li-white2009,baldry2012,bernardi2018} at $z=0$. These datasets provide a robust description of the SMF evolution. To build galaxy catalogs, we use cosmological N-body simulations that supply the merger trees hosting galaxies.

Three simulations are employed: YS50HR ($50\,\rm{Mpc/h}$ box), YS200 ($200\,\rm{Mpc/h}$), and YS300 ($300\,\rm{Mpc/h}$). All run with {\small GADGET4} \citep{springel2021} from $z=63$ to $z=0$ (100 snapshots between $z=20$ and 0), adopting Planck 2018 cosmology \citep{planck2020}. Their DM particle masses are $10^7$, $3.26 \times 10^8$, and $2.2 \times 10^9 \,M_{\odot}/h$, respectively (see \citealt{contini2023} for details).

Following C25, the free parameters are: the AGN efficiency $\kappa_{\rm AGN}$ (Eq.\ref{eqn:radiomode}), the scale velocity $V_{\rm scale}$ (Eq.\ref{eqn:ejection}), and the reincorporation factor $\gamma_2$.\footnote{$\dot{M}_{\rm ej} = \gamma_2 M_{\rm ej}/t_{\rm reinc}$, where $M_{\rm ej}$ is the gas in the ejecta reservoir and $t_{\rm reinc}$ the reincorporation timescale. See C24 for details.} The calibration for \texttt{AGNeject1} was already presented in C25, so here we focus on \texttt{AGNeject2}. The best-fit parameters are: $\kappa_{\rm AGN} = 2 \times 10^{-5}$ ($5 \times 10^{-5}$) $M_{\odot}$/yr, $V_{\rm scale}=500$ (415) km/s, and $\gamma_2=0.378$ (0.367) for \texttt{AGNeject1} (\texttt{AGNeject2}).

Finally, the models are run on merger trees from all three simulations to construct the catalogs. Due to different resolutions, we select halos above $\log M_{200}=11.0$ (YS50HR), 12.0 (YS200), and 14.0 (YS300). These combined subcatalogs form our final dataset for analysis.

\subsection{NewCluster}

\texttt{NewCluster} simulation (\citealt{han2025}) is a high-resolution cosmological zoom-in run designed to study galaxy formation in a massive cluster environment. It uses the \texttt{RAMSES-yOMP} code \citep{han2025a}, a hybrid-parallelized version of the RAMSES hydrodynamics solver \citep{teyssier2002}, and focuses on the evolution of a region centered on a galaxy cluster with a virial mass of approximately $5 \times 10^{14}\,M_\odot$ at redshift $z = 0$. The simulation starts from a periodic box of 100 Mpc$/h$ and zooms in on a region of radius $\sim 17.7$ Mpc$/h$, corresponding to $3.5\,R_{\rm{vir}}$.

In the zoom-in region, the mass resolution is $m_{\rm{DM}} = 1.6 \times 10^6 \,M_\odot$ for dark matter and $m_* = 2 \times 10^4 \,M_\odot$ for stellar particles. The spatial resolution ranges from 53 to 107 parsecs, depending on the expansion factor. The adopted cosmology follows WMAP-7 parameters: $H_0 = 70.3 \,\rm{km\,s^{-1}\,Mpc^{-1}}$, $\Omega_m = 0.272$, $\Omega_\Lambda = 0.728$, $\sigma_8 = 0.810$, and $n_s = 0.967$ \citep{komatsu2011}.

Radiative processes assume collisional ionization equilibrium, with metal-line and dust cooling allowing gas to cool below $10^4 ,\rm{K}$. A uniform UV background is switched on at $z=10$, with density-based self-shielding.

Star formation occurs in gas above $n_{\rm H} > 5 ,\rm{cm^{-3}}$, following a thermo-turbulent efficiency model. The simulation tracks nine elements (H, D, C, N, O, Mg, Fe, Si, S), adopting a \citet{chabrier2003} IMF. Yields come from SN II \citep{kobayashi2006}, SN Ia \citep{iwamoto1999}, and stellar winds \citep{leitherer1999}. Each SN Ia injects $10^{51}$ erg, with feedback implemented via the mechanical model of \citet{kimm2014}.

AGN feedback follows the dual-mode scheme of \citet{dubois2012}: jet mode (kinetic) for $f_{\rm Edd}<0.01$, and quasar mode (thermal) for $f_{\rm Edd}>0.01$, capped at unity. Further details are given in \citet{han2025}.

\section{Results and Discussion}
\label{sec:results}

In the following sections, we use our catalogs to carry out three main analysis: (1) to investigate the baryonic content of dark matter halos, (2) to quantify the relative contributions of SN and AGN feedback in ejecting hot gas beyond the virial radius, and (3) to compare our predictions with hydrodynamic simulations and observational estimates. Predictions from our models, including those from \texttt{NewCluster}, will be compared with results from other simulations as well as a wide range of observational data, followed by a detailed discussion of the implications. We emphasize that, unless otherwise stated, all model predictions are presented as median values, with the 16th and 84th percentiles shown when possible.

\subsection{The Baryonic Content in DM Halos}
\label{sec:baryon}

\begin{figure*}[t!]
\centering
\includegraphics[width=0.95\textwidth]{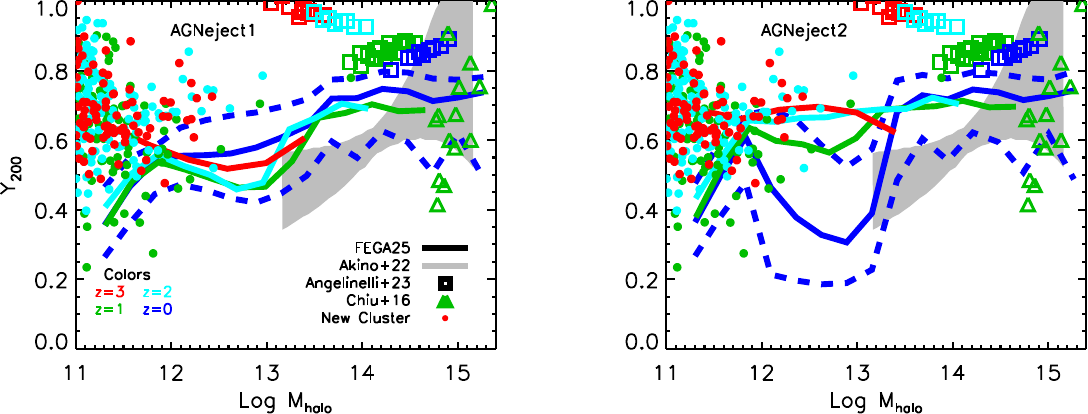}
\caption{Baryon fraction normalized to the universal value ($\rm{Y}{200}$) as a function of halo mass for the \texttt{AGNeject1} (left) and \texttt{AGNeject2} (right) models. Solid lines indicate different redshifts, compared with predictions from \texttt{NewCluster} (filled circles) and literature data (symbols in the legend). \texttt{AGNeject1} shows almost no evolution with time, while \texttt{AGNeject2} predicts a strong late-time decline in baryon content between $11.8 < \log M{\rm{halo}} < 13.4$. On small scales, \texttt{NewCluster} supports the near-constant trend of \texttt{AGNeject1}. On cluster scales, both models agree with \citet{chiu2016} and \citet{akino2022}, but fall below \citet{angelinelli2023}. The most distinctive feature is the deep $z=0$ cavity in \texttt{AGNeject2}, signaling enhanced baryon depletion at late times.}
\label{fig:fig1}
\end{figure*}

We begin our analysis in Figure~\ref{fig:fig1}, which shows the baryon fraction—normalized to the universal value ($\rm{Y}_{200}$)—as a function of halo mass for the \texttt{AGNeject1} (left panel) and \texttt{AGNeject2} (right panel) models. Solid lines represent model predictions at different redshifts, while filled circles correspond to results from our \texttt{NewCluster} simulation. We also include observational estimates from \citet{chiu2016} at $z\sim 0.9$, and \citet{akino2022} at $0<z<1$, along with predictions from the {\small MAGNETICUM} simulation \citep{dolag2016} by \citet{angelinelli2023}, with symbols as indicated in the legend.

A caveat should be noted: while our predictions are based on $\rm{Y}_{200}$ and $M_{200}$, both the observational and simulation data used for comparison refer to $\rm{Y}_{500}$. Since the baryon content within $R_{500}$ is necessarily smaller than within $R_{200}$, these external datasets should be interpreted as lower limits.

Focusing first on the comparison between our models, \texttt{AGNeject1} shows little redshift evolution, with only mild variation in a order of magnitude centered around $\log M_{\rm{halo}} \sim 12.7–12.8$. At lower halo masses, the predictions from \texttt{NewCluster} closely follow the nearly constant trend suggested by \texttt{AGNeject1}. At $z=0$, the baryon fraction increases from about 0.4 at $\log M_{\rm{halo}} \sim 11.2$ to roughly 0.75 for the most massive halos, reflecting a clear correlation with halo mass.

In contrast, \texttt{AGNeject2} exhibits a more pronounced redshift dependence over the mass range $11.8 < \log M_{\rm{halo}} < 13.4$, especially between $z=1$ and the present. This results in a distinct ``cavity" in the baryon fraction at $z=0$ within that range. Outside this interval, the two models predict similar baryon fractions, indicating that the differences arise primarily at intermediate halo masses and late times—consistent with a more efficient, redshift-dependent removal of baryons in \texttt{AGNeject2}.

Compared to {\small MAGNETICUM}, both models predict lower baryon fractions—expected given the different apertures—but they are broadly consistent with observational data from \citet{chiu2016} and \citet{akino2022} for halos above $\log M_{\rm{halo}} \sim 13$, even accounting for the likely upward correction needed in the observed values.

\begin{figure*}[t!]
\centering
\includegraphics[width=0.95\textwidth]{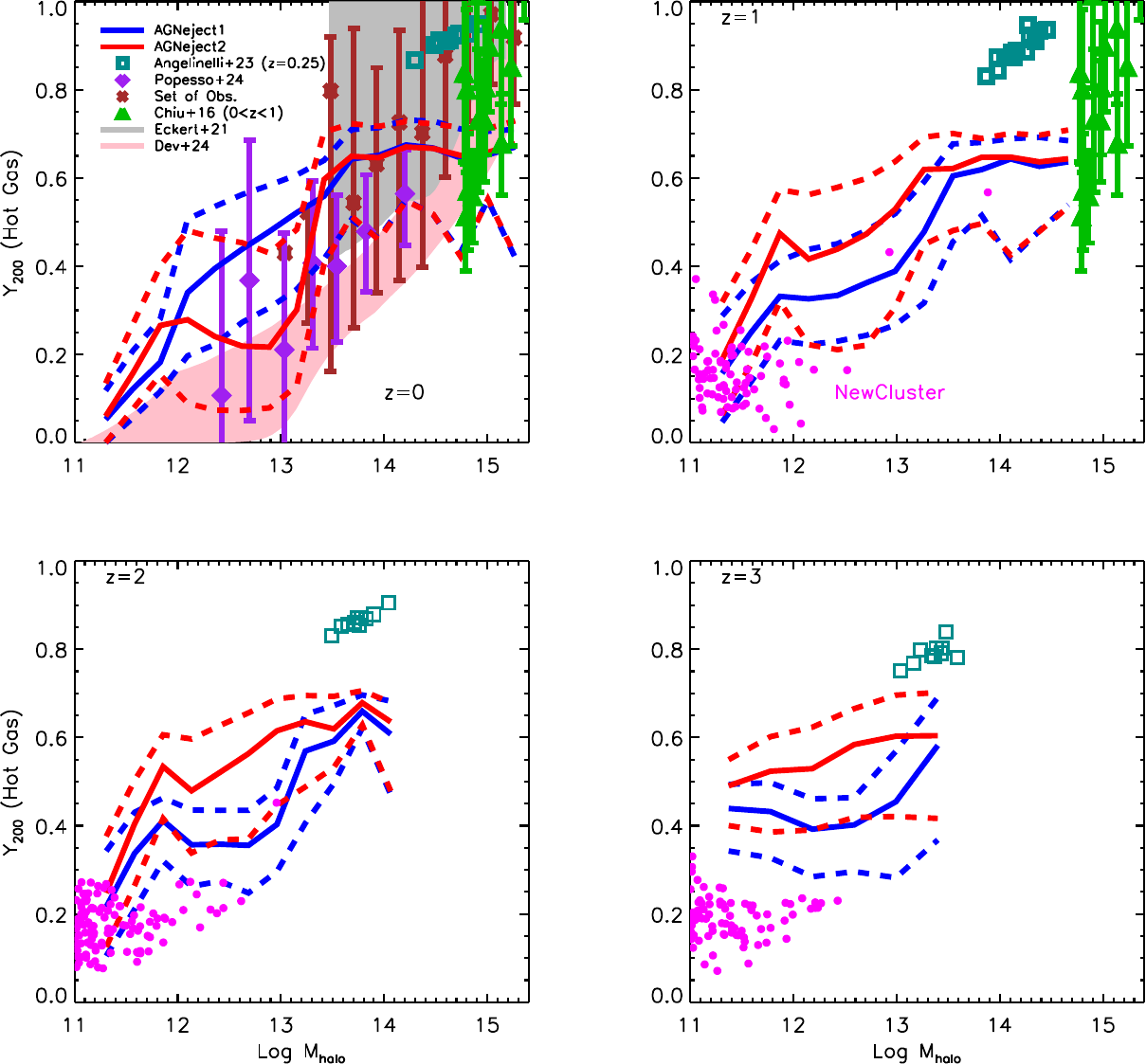}
\caption{Hot gas fraction normalized to the universal baryon fraction as a function of halo mass at different redshifts (separate panels), predicted by \texttt{AGNeject1} (blue) and \texttt{AGNeject2} (red). Results are compared with observational constraints and simulations at $z=0$ and $z=1$, and with simulations only at higher redshifts. Predictions from \texttt{NewCluster} are also shown at $z \geq 1$ (magenta circles). Both models broadly agree with observations at $z=0$ and show hints of convergence at $z=1$, though they predict lower fractions than {\small MAGNETICUM} \citep{angelinelli2023} at all epochs. The main difference arises in Milky Way–like halos: \texttt{AGNeject1} is more effective until $z=1$, while \texttt{AGNeject2} becomes dominant in the last $\sim 7$ Gyr. At $z=3$, \texttt{NewCluster} gives slightly lower values but follows the same trends. This explains the deep baryon fraction cavity at $z=0$ in \texttt{AGNeject2} (Figure~\ref{fig:fig1}).} 
\label{fig:fig2}
\end{figure*}

Baryons in our models are distributed among four main components: stars (including both those bound to galaxies and the intracluster light), cold and hot gas within the virial radius, and ejected gas—namely, hot gas residing outside the virialized region. As shown in Figure~\ref{fig:fig1}, the total baryon fraction spans values between approximately 0.4 and 0.75, indicating that a substantial fraction of baryons is located beyond the virial boundary.

To investigate the gaseous component in more detail, Figure~\ref{fig:fig2} displays the normalized hot gas fraction as a function of halo mass, with predictions from \texttt{AGNeject1} (blue lines) and \texttt{AGNeject2} (red lines) across different redshifts (each shown in separate panels). These are compared to predictions from the \texttt{NewCluster} simulation and a compilation of observational data from the literature: brown stars from \citet{vikhlinin2006}, \citet{arnaud2007}, \citet{sun2009}, \citet{rasmussen2009}, \citet{pratt2009}, \citet{mahdavi2013}, \citet{gonzalez2013}, \citet{lovisari2015}, \citet{eckert2016}, \citet{pearson2017}, \citet{eckert2019}, \citet{mulroy2019}, and \citet{ragagnin2022}; magenta diamonds from \citet{popesso2024}; green diamonds from \citet{chiu2016}; a grey shaded region from \citet{eckert2021}; a pink shaded region from \citet{dev2024}; and predictions from the {\small MAGNETICUM} simulation.

Before delving into the results, we note an important caveat. As in Figure~\ref{fig:fig1}, many of the observational estimates and {\small MAGNETICUM} predictions refer to gas fractions measured within $R_{500}$, rather than $R_{200}$. Unlike the case of total baryons, however, a correction can be applied to translate $\rm{Y}_{500}$ values into their $\rm{Y}_{200}$ equivalents. To perform this conversion, we used the empirical relations provided by \citet{popesso2024}, which link $\rm{Y}_{200}$ to $R_{200}$ and $\rm{Y}_{500}$ to $R_{500}$, based on fits to their data. This allows us to compare our model predictions directly to the corrected observational estimates shown in Figure~\ref{fig:fig2}.

Focusing first on $z=0$, where observational constraints are most abundant, we find that the difference between the two AGN feedback models—previously identified as a cavity in Figure~\ref{fig:fig1}—is entirely attributable to a more pronounced ejection of hot gas in \texttt{AGNeject2} relative to \texttt{AGNeject1}. Outside of the intermediate mass range affected by this process, both models predict nearly identical hot gas fractions, consistent with their similar total baryon fractions. Importantly, despite the intrinsic scatter in the data, both models agree well with observations over a wide halo mass range, from $\log M_{\rm{halo}} \sim 12.5$ up to the most massive clusters. Once again, the {\small MAGNETICUM} simulation tends to overpredict the hot gas content relative to our models.

It is worth noting that, although not displayed in our figures, several recent observational studies suggest that Milky Way–mass halos may host relatively high baryon fractions. For example, \citet{das2023}, using the thermal SZ effect, reported a non-monotonic dependence of the baryon fraction on galaxy mass, with a peak around galaxies of $\log M_* \sim 11$. Similarly, \citet{nicastro2023} and \citet{mathur2023}, through X-ray absorption measurements, estimated $\log M_{\rm CGM} \sim 11$ for halos of $\log M \sim 12$, corresponding to a normalized hot gas fraction of roughly 0.6. Taken together, these studies support the presence of a declining baryon fraction with increasing halo mass in the range $\log M \sim 12-13$, consistent with the cavity predicted by \texttt{AGNeject2}. This agreement with independent observations reinforces the need for feedback mechanisms, such as AGN-driven hot gas ejection, in order to reproduce the full baryon cycle in galaxies and halos.

\begin{figure*}[t!]
\centering
\includegraphics[width=0.95\textwidth]{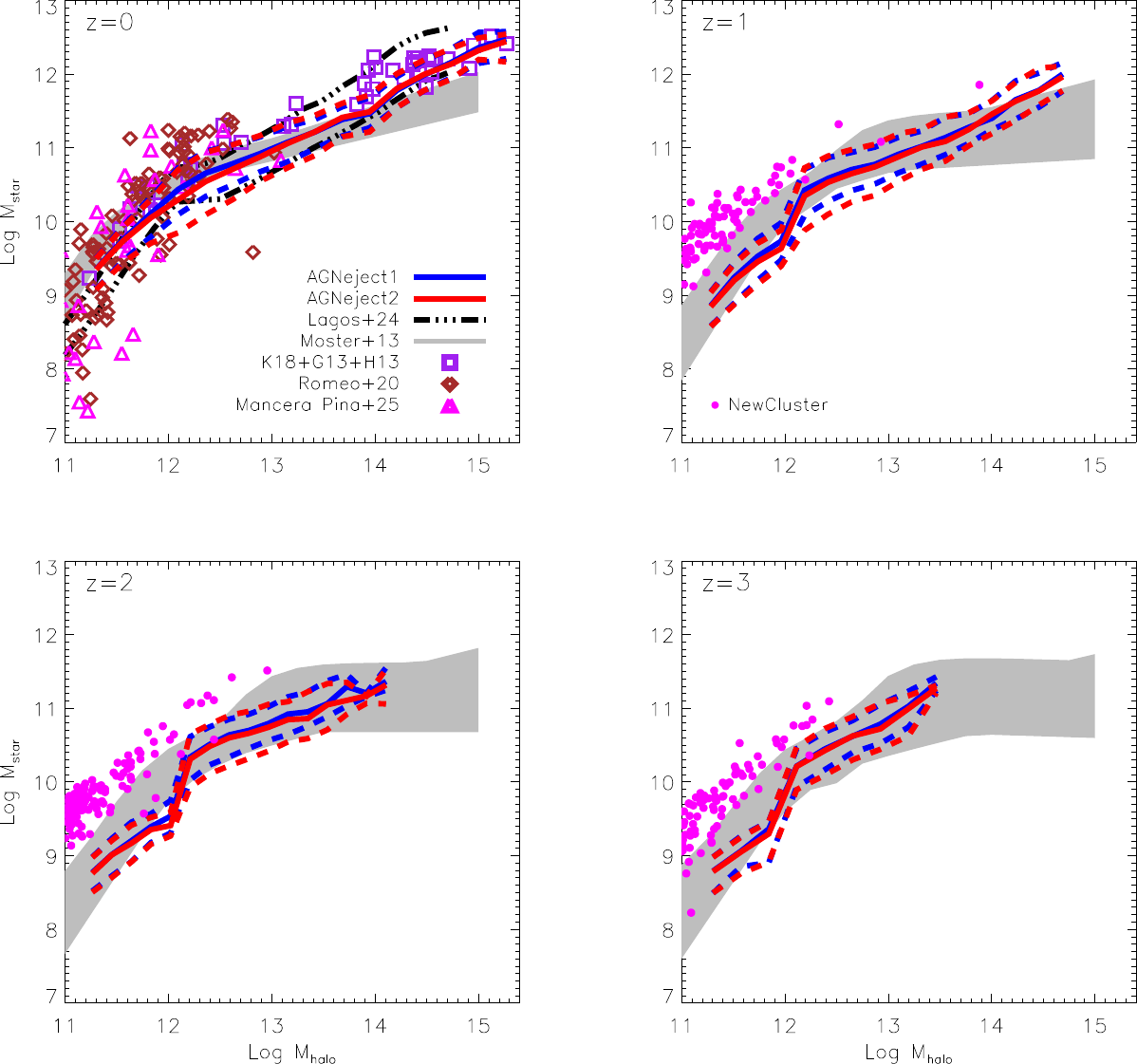}
\caption{Stellar-to-halo mass relation for central galaxies at different redshifts (panels), as predicted by \texttt{AGNeject1} (blue) and \texttt{AGNeject2} (red). At $z=0$, predictions are compared with the SHARK2 SAM \citep{lagos2024}, the empirical relation of \citet{moster2013}, and observational estimates from \citet{romeo2020}, \citet{hinshaw2013}, \citet{gonzalez2013}, \citet{kravtsov2018}, and \citet{mancera-pina2025}. At $z \geq 1$, results are compared with \citet{moster2013} and the \texttt{NewCluster} simulation. Both models agree well with semi-analytic, empirical, and observational constraints at $z=0$. At higher redshifts, \texttt{NewCluster} tends to overpredict stellar masses in low-mass halos. Matching this relation over cosmic time is crucial, especially in low-mass halos where the central galaxy contains most of the stellar mass within the virial radius.}
\label{fig:fig3}
\end{figure*}

At higher redshift, in particular at $z=1$, we start to see the difference between our models: \texttt{AGNeject1} predicts a lower hot gas fraction than \texttt{AGNeject2} in the mass range corresponding to the cavity at the present time. This is because \texttt{AGNeject1} is effective from early times down to $z=0$, whereas \texttt{AGNeject2} becomes highly effective only after $z \sim 1$. However, given the constancy of the hot gas fraction on cluster scales, our predictions remain consistent with the observations by \citet{chiu2016}, although they do not match those from {\small MAGNETICUM}, which show a constant trend even at higher redshift (see bottom panels). \texttt{NewCluster}, on the other hand, aligns well with our models at small scales up to $z=2$, but this agreement breaks down at $z=3$.

Comparing our results qualitatively with earlier studies—specifically C25 and \citet{popesso2024}—we find similar conclusions. These authors emphasized the necessity, in numerical models, of a feedback mechanism capable of ejecting hot gas from halos without significantly altering the stellar and cold gas components. This is precisely what our models achieve, as the next plot will illustrate. Nonetheless, it remains challenging to distinguish between the two models due to the scarcity of observational data at $z > 1$ in the halo mass range where the cavity begins to develop—a feature entirely driven by more efficient hot gas expulsion.

One final point worth noting is that the normalized hot gas fraction at $z=0$—where observational coverage is most comprehensive—rises from roughly 0.3 at $\log M_{\rm{halo}} \sim 12$ to about 0.65 in the most massive clusters. Comparing these figures with the total baryon fractions (which include cold gas and stars), we infer that, on cluster scales, approximately 92\%-93\% of the baryons reside in the hot phase, with this percentage declining toward lower halo masses. As previously shown in C25, on smaller scales, both the stellar and cold gas components become more significant.

We conclude this section by showing that our models also reproduce the stellar content of halos. Figure~\ref{fig:fig3} presents the stellar-to-halo mass relation for central galaxies at different redshifts, with our models color-coded as before. At $z=0$, they match the SHARK2 SAM \citep{lagos2024}, the empirical relation of \citet{moster2013}, and observational data from \citet{romeo2020}, \citet{hinshaw2013}, \citet{gonzalez2013}, \citet{kravtsov2018}, and \citet{mancera-pina2025}. Across the full mass range, both models agree well with data and show no visible differences.

At higher redshifts, our models also follow the \citet{moster2013} relation up to $z=3$, confirming their ability to reproduce the stellar mass of central galaxies at all times. By contrast, \texttt{NewCluster} overpredicts stellar masses by about an order of magnitude, an issue to be explored in Seo C., et al. (in prep).

These results demonstrate that our SAM reliably captures the stellar-to-halo mass relation at all epochs. Since stars are one of the main components in the baryon fraction, this confirms that the models describe them accurately. Given that tensions between model and data for baryon and hot gas fractions arise mostly on group scales—where cold gas is negligible and stellar mass is a good tracer—the discrepancies must be driven by how much hot gas feedback expels.

This conclusion aligns with C25 and \citet{popesso2024}: the physical mechanisms removing hot gas must leave the stellar and cold components unaffected. Both of our models achieve this in different ways and with varying strength over time. With this established, we now turn to the two feedback channels able to expel hot gas—SN and AGN—to assess their relative roles across mass and redshift.

\subsection{Roles of SN and AGN feedback}
\label{sec:SNAGN}

\begin{figure*}[t!]
\centering
\includegraphics[width=0.95\textwidth]{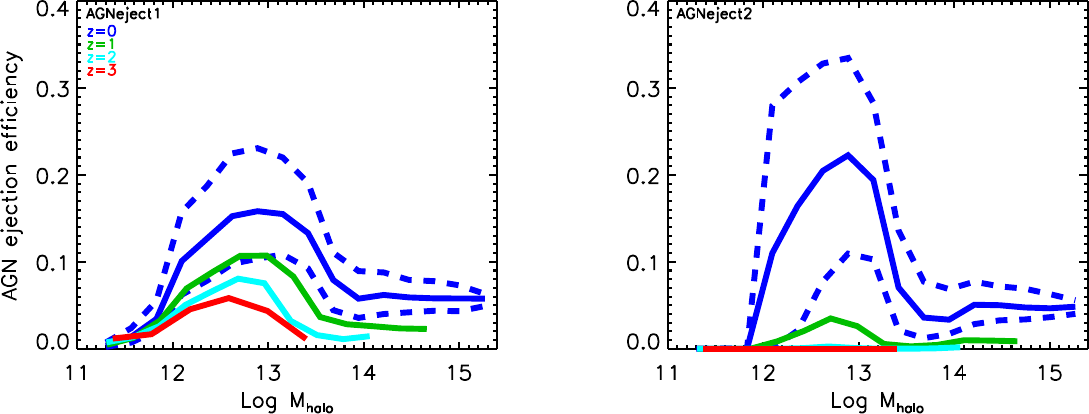}
\caption{Efficiency of AGN feedback in ejecting gas beyond the virial radius as a function of halo mass. The efficiency is defined as the total gas mass expelled by AGN up to a given redshift (including all progenitors), normalized by the halo mass at that epoch. Predictions from our two models are shown: \texttt{AGNeject1} (left) and \texttt{AGNeject2} (right), with colors indicating different redshifts. In \texttt{AGNeject1}, efficiency gradually rises with time, peaking between $12.5 < \log M_{\rm{halo}} < 13$, depending on redshift, and flattening to $\sim 5\%$ on cluster scales. By contrast, \texttt{AGNeject2} remains negligible until $z=1$, then increases sharply at $\log M_{\rm{halo}} > 12$, reaching a peak near $\log M_{\rm{halo}} \sim 13$ at $z=0$. This highlights the contrasting timing and strength of AGN-driven gas ejection in the two models.}
\label{fig:fig4}
\end{figure*}

\begin{figure*}[t!]
\centering
\includegraphics[width=0.95\textwidth]{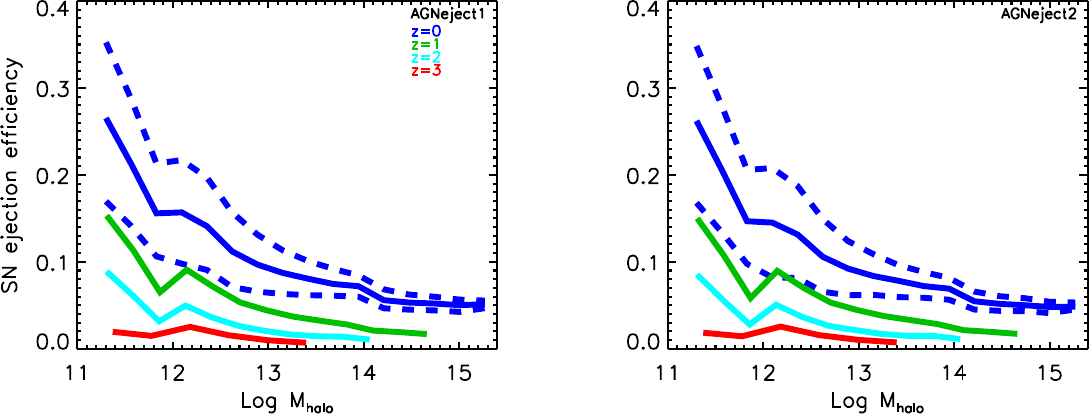}
\caption{SN feedback efficiency as a function of halo mass for the \texttt{AGNeject1} (left) and \texttt{AGNeject2} (right) models, shown at different redshifts (color-coded lines as indicated in the legend), following the same format as Figure \ref{fig:fig4}. SN feedback is most effective in low-mass halos and progressively declines with increasing halo mass. The two panels are identical, reflecting the fact that the AGN feedback implementations in these models do not influence the SN feedback.}
\label{fig:fig5}
\end{figure*}

\begin{figure*}[t!]
\centering
\includegraphics[width=0.95\textwidth]{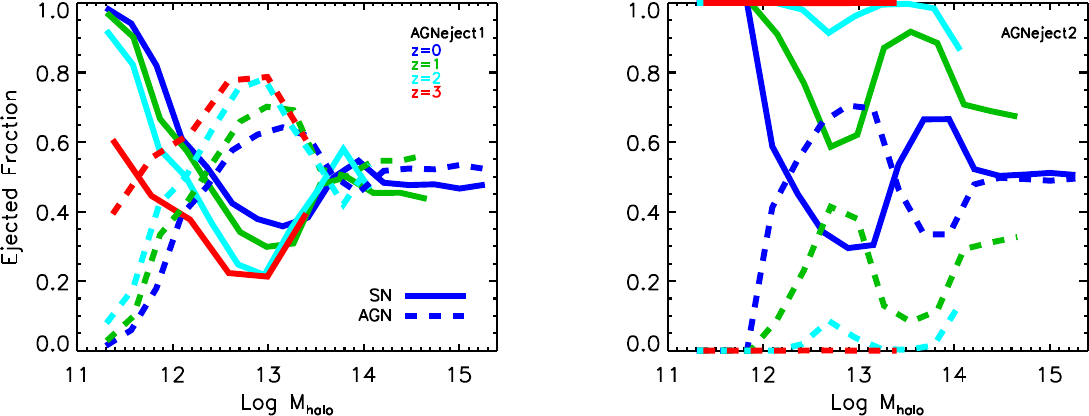}
\caption{Fraction of gas ejected beyond the virial radius by SN (solid lines) and AGN (dashed lines) feedback, as a function of halo mass and at various redshifts (indicated by different colors), for the \texttt{AGNeject1} (left panel) and \texttt{AGNeject2} (right panel) models. The fractions shown represent the contribution of each feedback mode relative to the total ejected gas (i.e., SN plus AGN). The figure highlights the marked differences between the two models. In \texttt{AGNeject1}, AGN feedback dominates over SN feedback in the halo mass range $11.5 < \log M_{\rm{halo}} < 13.5$, depending on redshift, with the dominance centered at $\log M_{\rm{halo}} \sim 13$ at $z = 0$. In contrast, \texttt{AGNeject2} shows negligible AGN ejection until $z \leq 1$, after which it becomes efficient and overtakes SN feedback in the range $12.3 < \log M_{\rm{halo}} < 13.2$ at $z = 0$. At the high-mass end ($\log M_{\rm{halo}} > 14$), SN and AGN feedback contribute comparably in both models.}
\label{fig:fig6}
\end{figure*}

As mentioned several times above, SN feedback and, since the implementation of the hot gas ejection mode in C25, AGN feedback are the only mechanisms capable of expelling gas from the virialized region of halos. A key objective is to understand the level of gas expulsion associated with each mechanism—that is, how efficient they are over time and across different halo masses.

To address this, in Figure~\ref{fig:fig4} we plot the efficiency of AGN feedback as a function of halo mass for \texttt{AGNeject1} (left panel) and \texttt{AGNeject2} (right panel), at various redshifts as indicated by the color legend. We define the efficiency as the total amount of gas expelled by AGN feedback up to the redshift in question, accounting for all progenitors in the merger trees.

\texttt{AGNeject1} exhibits a gradual increase in efficiency from high to low redshift, peaking at $z=0$ with an efficiency of about 17\% at $\log M_{\rm{halo}} \sim 13$. On either side of this characteristic mass, the efficiency decreases moderately, stabilizing at approximately 5\% for $\log M_{\rm{halo}} > 14$. In contrast, \texttt{AGNeject2} shows markedly different behavior. Although its efficiency profile with halo mass resembles that of \texttt{AGNeject1} at $z=0$, the peak is higher (around 23\%), and the efficiency at higher redshifts ($z > 1$) is negligible across all halo masses.

This divergence emphasizes that \texttt{AGNeject1} gradually expels hot gas over time, while \texttt{AGNeject2} remains largely inactive for about 7 Gyr and then becomes suddenly effective, even surpassing the performance of \texttt{AGNeject1} between $0 < z < 1$. In summary, by $z=0$, both models are most effective within the range $12 < \log M_{\rm{halo}} < 14$, with an efficiency of roughly 5\% on cluster scales.

In Figure~\ref{fig:fig5}, we present the counterpart of Figure~\ref{fig:fig4} for SN feedback. At first glance, it is evident that both models behave identically in this case, indicating that AGN feedback does not influence the behavior of SN feedback. The efficiency of SN feedback decreases with halo mass and increases with redshift. The smallest halos in our sample experience the strongest SN feedback, with efficiencies reaching up to 30\%, while even on cluster scales the efficiency remains non-negligible at around 5\%. It is worth emphasizing that on cluster scales, the efficiencies of SN and AGN feedback are comparable—an important and non-trivial result.

To directly compare SN and AGN feedback within the same framework, in Figure~\ref{fig:fig6} we plot their respective contributions to the total gas expelled as a function of halo mass and redshift \footnote{In our model only SN and AGN can eject gas.}. SN feedback is represented by solid lines, and AGN feedback by dashed lines, with different colors indicating redshift. The left and right panels correspond to the \texttt{AGNeject1} and \texttt{AGNeject2} models, respectively. The two models display clearly distinct behaviors: \texttt{AGNeject1} shows a weak dependence on redshift, whereas \texttt{AGNeject2} exhibits a strong redshift dependence, as expected from its delayed onset of AGN activity.

Notably, in both models, SN feedback dominates the total gas expulsion except within the characteristic mass ranges where AGN feedback becomes particularly efficient. Specifically, AGN feedback dominates in halos with $12.5 < \log M_{\rm{halo}} < 13.5$ for \texttt{AGNeject1}, and $12.2 < \log M_{\rm{halo}} < 13.3$ for \texttt{AGNeject2}, both at $z=0$. Thus, SN feedback is the primary driver of gas ejection in low-mass and Milky Way-like halos, while AGN feedback takes over in the subsequent mass decade, with both mechanisms contributing roughly equally on large group and cluster scales. It is important to note that in our SAM only SN and AGN feedback are capable of ejecting gas. On cluster scales, their contributions are comparable, although the absolute amount of expelled gas can be relatively small. We also remind the reader that the plot accounts for all progenitors in the merger trees, not just the main branch.

Our findings are qualitatively consistent with those of \citet{ayromlou2023} (but see also the analysis in \citealt{wright2024}), who analyzed three sets of cosmological hydrodynamic simulations—EAGLE \citep{schaye2015}, IllustrisTNG \citep{pillepich2018}, and SIMBA \citep{dave2019}. These authors showed that feedback processes play a crucial role in redistributing gas within and beyond halo boundaries, reducing the baryon fraction inside halos and transferring hot gas outside the virial radius.

A particularly interesting result from their study is the concept of the closure radius—the radius at which the enclosed baryon fraction equals the universal value. They found that the closure radius decreases with increasing halo mass. For instance, in low-mass halos with $\log M_{\rm{halo}} = 11$, this radius can be as large as 5–10 times the virial radius, depending on the simulation. In contrast, for massive groups and clusters, the closure radius lies within the virial radius itself.

While our models predict baryon fractions of about 75\% in massive halos—somewhat lower than the closure point identified in the simulations—there is strong agreement regarding the mass dependence of feedback efficiency: both SN and AGN feedback are found to be critical in similar halo mass ranges. 

To conclude this section, we now turn to a detailed comparison between the baryon fractions predicted by our models and those reported by \citet{ayromlou2023} in their analysis.

\subsection{\texttt{FEGA25} vs Simulations}
\label{sec:numcomp}

\begin{figure*}[t!]
\centering
\includegraphics[width=0.95\textwidth]{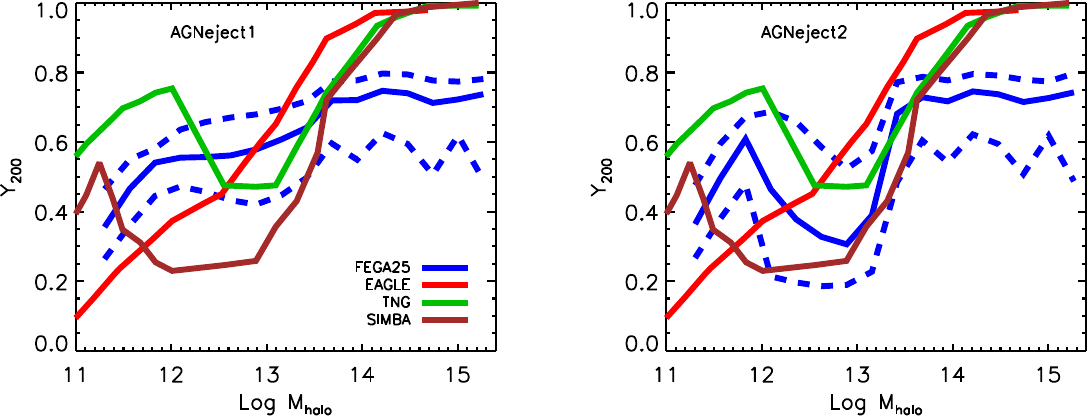} 
\caption{Baryon fraction normalized to the universal value as a function of halo mass at $z=0$, as predicted by \texttt{AGNeject1} (left) and \texttt{AGNeject2} (right). Predictions are compared with the EAGLE (\citealt{schaye2015}, red), IllustrisTNG (\citealt{pillepich2018}, green), and SIMBA (\citealt{dave2019}, brown) simulations. Neither model matches the simulations across the full halo mass range, but \texttt{AGNeject2} resembles IllustrisTNG and SIMBA by producing a distinct ``cavity'' in the baryon fraction, though with different depth and location. Unlike the simulations, which rise toward the universal fraction on cluster scales, our models still predict a $\sim$25\% baryon deficit, implying a substantial reservoir of baryons remains beyond the virialized region.}
\label{fig:fig7}
\end{figure*}

We end our analysis with a full comparison between our models and predictions from hydrodynamic simulations. \citet{ayromlou2023} convincingly demonstrated that the EAGLE, IllustrisTNG, and SIMBA simulations all predict different baryon fractions as a function of halo mass, emphasizing that the baryonic content of halos is highly sensitive to the specific physical processes implemented in each model.

In Figure~\ref{fig:fig7}, we show the baryon fraction—normalized to the universal value—as a function of halo mass at $z=0$, as predicted by our models: \texttt{AGNeject1} (left panel) and \texttt{AGNeject2} (right panel). For comparison, we also overplot the predictions from the above-mentioned simulations, with colors indicated in the legend. It is evident that both the simulations and our models predict significantly different baryon fractions and trends. Nonetheless, some common features can be identified across the simulations. First, they all converge on cluster scales ($\log M_{\rm{halo}} > 14$), where the baryon fraction approaches the universal value. Second, while EAGLE exhibits a monotonic increase in baryon fraction from low to high halo masses, both IllustrisTNG and SIMBA show a pronounced dip—that we named cavity—in the intermediate-mass range, a feature also present in our model \texttt{AGNeject2}. However, the location, width, and depth of this cavity vary substantially among the simulations.

From this comparison, we can conclude that \texttt{AGNeject1} aligns more closely with EAGLE in terms of the general trend, particularly the absence of a cavity. Conversely, \texttt{AGNeject2}, especially for $\log M_{\rm{halo}} < 13.7$, resembles the behavior seen in both IllustrisTNG and SIMBA. Given the observational comparisons shown in Figures~\ref{fig:fig1} and \ref{fig:fig2}, and especially considering the large scatter in the data, we conclude that it is currently not possible to rule out any of the models or the physical assumptions implemented in them. To determine whether the cavity is a genuine feature, more observational data are needed for halos with $\log M_{\rm{halo}} < 12$, including accurate measurements of both baryon and hot gas fractions, and with reduced observational scatter.

This feature of the rapid development of the cavity in halos with $12 \lesssim \log M_{\rm halo} \lesssim 13$ between $z=1$ and the present time has a physical interpretation. In our \texttt{AGNeject2} model, this late-time behavior arises because the hot gas ejection mode remains inefficient at high redshift, when the supply of cosmological accretion is still strong and AGN feedback energy is largely spent counteracting cooling flows. After $z \sim 1$, however, two conditions change simultaneously: (i) the accretion of fresh gas slows down as halos transition toward quasi-hydrostatic equilibrium, and (ii) BHs have grown sufficiently massive that the cumulative AGN energy input surpasses the binding energy of the hot atmospheres. This means that, in this time and halo mass regimes, the excess energy is no longer required to offset cooling but can instead drive substantial amounts of gas beyond the virial radius.

This transition explains why \texttt{AGNeject2} is almost inactive for the first $\sim 7$ Gyr of cosmic time, yet becomes suddenly powerful in expelling hot gas at later epochs. The contrast with \texttt{AGNeject1}, which steadily ejects gas throughout cosmic history, highlights that the two implementations probe different physical regimes: a gradual, continuous removal versus a delayed but more explosive release once AGN feedback overtakes the capacity of halos to retain gas. As a result, \texttt{AGNeject2} naturally produces the fast depletion of gas, a trend that qualitatively matches the cavity hinted by some observational studies and hydrodynamic simulations, as seen in the analysis above.

A full comparison between observations and simulations was recently presented by \citet{popesso2024}, who used the public eROSITA Early Data Release event file of the eFEDS field \citep{brunner2022} to compare the hot gas fraction within $R_{500}$ with several simulations: BAHAMAS \citep{salcido2023}, FLAMINGO \citep{schaye2023}, IllustrisTNG, MAGNETICUM, SIMBA, and MillenniumTNG \citep{pakmor2023}. They found that IllustrisTNG and MillenniumTNG generally overpredict hot gas fractions, while FLAMINGO and BAHAMAS—with similar AGN feedback—are closer to the data. Only MAGNETICUM reproduces the observed relation very well, with SIMBA consistent within $1\sigma$. Interestingly, MAGNETICUM, the best match to the data, does not display a cavity as seen in some simulations and in our model \texttt{AGNeject2}, but instead follows a power-law trend similar to that observed.

Overall, the comparison shows a wide variety of behaviors. Some simulations, as well as our model \texttt{AGNeject2}, predict a cavity in baryon and hot gas fractions, while others—including our model \texttt{AGNeject1}—do not. Current observations appear to favor a power-law behavior, but probing halos below $\log M_{\rm{halo}} \sim 12$ remains crucial before drawing firm conclusions on the physical models used in simulations and SAMs.

In light of these findings, we present our final conclusions in the next section.

\section{Conclusions}
\label{sec:conclusion}

In our state-of-the-art semi-analytic model, \texttt{FEGA25}, we have implemented an additional mechanism—alongside SN feedback—for ejecting hot gas beyond the virial radius of halos. This mechanism is directly associated with AGN feedback and is realized in two different forms: the \texttt{AGNeject1} model, identical to the one introduced in \citet{contini2025a}, and the \texttt{AGNeject2} model, which differs in strength and becomes primarily active at $z<1$.

Our models predict baryon and hot gas fractions as a function of halo mass in broad agreement with current observational constraints. Notably, while \texttt{AGNeject1} yields a monotonic increase in both fractions with halo mass, \texttt{AGNeject2} predicts a distinct cavity at halo masses below $\log M_{\rm{halo}} \sim 13$, a feature commonly seen in several hydrodynamic simulations. Furthermore, \texttt{AGNeject2} exhibits a redshift dependence in the mass range where the cavity appears, while \texttt{AGNeject1} shows minimal evolution. The primary difference between the two models lies in the time-dependent strength of AGN feedback; however, both are capable of efficiently ejecting hot gas from halos.

Both models successfully reproduce the stellar-to-halo mass relation up to $z=3$, and match observational data at $z=0$, which is a significant achievement given the broad dynamic range in both halo mass and redshift. This success is particularly relevant in the context of the hot gas content in galaxy groups, where the cold gas fraction becomes negligible. The implication is that any viable numerical method must include a mechanism that can remove hot gas without significantly altering the stellar and cold gas components—consistent with recent observational findings.

Our results also highlight the relative importance of SN and AGN feedback across different halo mass scales. In line with recent hydrodynamic simulation results at $z=0$—which we extend to higher redshifts—we find that SN feedback dominates for $\log M_{\rm{halo}} \lesssim 12.5$, while AGN feedback becomes increasingly significant at higher halo masses. On group and cluster scales, both feedback mechanisms contribute comparably. However, at higher redshifts, the models diverge: \texttt{AGNeject1} maintains a nearly constant division between SN and AGN feedback, whereas \texttt{AGNeject2} exhibits significant evolution, with SN feedback remaining dominant down to $z=1$, at all scales.

When compared to both simulations and observations, the two models show differing levels of agreement depending on the specific regime. \texttt{AGNeject1} is more consistent with observational data, as it predicts a smooth, monotonic increase in baryon and hot gas fractions with halo mass. In contrast, \texttt{AGNeject2} reproduces the cavity seen in several simulations. Nonetheless, we do not discard either model. The lack of observational data at small halo masses—particularly below $\log M_{\rm{halo}} \sim 12.5$—where \texttt{AGNeject2} and several simulations predict a re-rise in the gas fractions, coupled with the significant observational scatter at all mass scales, makes it premature to rule out any model. In order to assess the physical plausibility of the cavity, and thus constrain the most realistic feedback mechanism, more observational data (besides those quoted above) are needed below $\log M_{\rm{halo}} \sim 12.$–12.5, with significantly reduced scatter (ideally at least a third to half of the current levels).

Despite the differing predictions of the two models, one robust conclusion emerges: as also supported by observational evidence, SN feedback alone is insufficient to account for the ejection of hot gas. Several simulation-based studies, along with our current semi-analytic implementation, demonstrate that AGN feedback in the form of hot gas ejection is a viable and necessary mechanism. Therefore, it must be included—and ideally refined—in any numerical framework aiming to accurately model galaxy formation and the baryon cycle in halos.

An interesting extension of this work concerns the cosmological applications of baryon fraction studies. Accurate modeling of the baryon content in halos is crucial for cluster cosmology and has recently been linked to new probes such as fast radio bursts, where the circum galactic medium gas fraction enters the interpretation of dispersion measures \citep{zhang2025,liu2025}. Our SAM, with its flexible feedback prescriptions, can therefore provide predictions directly relevant for these emerging cosmological tests.


\section*{Acknowledgements}
The authors thank the referee for his very constructive comments which helped to improve the manuscript.
E.C. and S.K.Y. acknowledge support from the Korean National Research Foundation (RS-2025-00514475). E.C. and S.J. acknowledge support from the Korean National Research Foundation (RS-2023-00241934). All the authors are supported by the Korean National Research Foundation (RS-2022-NR070872). J.R. was supported by the KASI-Yonsei Postdoctoral Fellowship and by the Korea Astronomy and Space Science Institute under the R\&D program (Project No. 2023-1-830-00), supervised by the Ministry of Science and ICT. This work was also partially supported by the Institut de Physique des deux infinis of Sorbonne Université and by the ANR grant ANR-19-CE31-0017 of the French Agence Nationale de la Recherche.

\section*{Data Availability}
The data used in this work, from our semi-analytic model and NewCluster, are available in HARVARD Dataverse:
\dataset[doi: 10.7910/DVN/EGKUF1]{\doi{10.7910/DVN/EGKUF1}}

\bibliography{paper}{}
\bibliographystyle{aasjournalv7}



\end{document}